\documentclass[runningheads]{svmult}

\usepackage{makeidx}   
\usepackage{graphicx}  
\usepackage{subeqnar}  
\usepackage{multicol}  
\usepackage{physprbb}  
\makeindex             

\begin{document}

\newcommand {\pks} {{\small PKS}$\;2155$-304}
\newcommand {\mkn} {Mkn$\;421$}
\newcommand {\dg} {$^\circ$}
\newcommand {\hess} {{\small H$\cdot$E$\cdot$S$\cdot$S$\cdot$}}
\newcommand {\bighess} {{H$\cdot$E$\cdot$S$\cdot$S$\cdot$}}
\newcommand {\gradi} {^{\circ} }
\newcommand {\gamm} {$\gamma$}
\newcommand {\gr} {$\gamma$-ray}

%
\title*{
Results from Observations of AGNs with the \protect\newline
\bighess\ Telescope System and Future Plans
}
\toctitle{
Results from Observations of AGNs with the \protect\newline
\hess\ telescope system and Future Plans
}
%
%
\titlerunning{\hess\ AGN Results and Future Plans}
%
\author{Michael Punch 
for the \hess\ Collaboration
}
\authorrunning{Michael Punch
for the \hess\ Collaboration
}
%
%
\institute{
Physique Corpusculaire et Cosmologie, IN2P3/CNRS, Coll{\`e}ge de France,
\\
11 Place Marcelin Berthelot, F-75231 Paris Cedex 05, France
}

\maketitle              

\begin{abstract}

 The \hess\ (High Energy Stereoscopic System) Phase-I is comprised of
 four Imaging Atmospheric Cherenkov Telescopes (IACTs) for observation
 of galactic and cosmic sources of Very High Energy (VHE) gamma rays,
 with a significant improvement in sensitivity and a detection
 threshold below that of previous {\small IACT}s.  Observations of
 Active Galactic Nuclei (AGNs) since the start of operations in June
 2002 are presented, in particular for \pks\ and \mkn , along plans
 for Phase-II.

\end{abstract}

\section{The \bighess\ Telescope System}

The \hess\ detector for observation of $>100\;\rm GeV$ \gr s has been
operating since June, 2002 in the Khomas highlands of Namibia (23\dg
S, 15\dg E, $1.8\;\rm km$ a.s.l.).  It captures the Cherenkov light
emitted by cascades of particles in the atmosphere initiated by a
\gr\ or charged cosmic ray incident on the atmosphere. The
Cherenkov pulses ($\lambda\sim 350\;\rm nm$) are brief (few ns),
faint, and illuminate a light-pool of diameter $\sim 250\;\rm m$ on
the ground for vertical cascades.  The Cherenkov images of these
cascades, roughly cometary in shape with an angular extent of a few
mrad, can be seen by a detector anywhere in the light-pool equipped
with a sufficiently fast and sensitive camera. This permits the
estimation of the nature of the initiating particle (signal
\gr\ or background cosmic-ray) and the measurement of its angular
origin and energy.  The Atmospheric Cherenkov technique intrinsically
has a large ($\sim 50000\;\rm m^2$) collection area, though with a
small field of view (few degrees). Observations must take place on
clear, moonless nights.

The detector, in its Phase-I, consists of four IACTs in a square of
side $120\;\rm m$.  Each telescope mount has a tessellated mirror of
$107\;{\rm m}^2$ area 
with a camera in the focal plane at $15\;{\rm m}$.  The camera
contains 960 photo-multipliers (PMs) with 0.16\dg\ pixel-size,
5\dg\ field of view.  The read-out electronics, all contained
within the camera, is triggered when the signal from a number of PMs
exceeds a trigger threshold in an effective $\sim 1.3\;\rm ns$ trigger
window.  The PM signals, which are stored in an analogue
memory while awaiting the trigger, are then read out, digitized, and
integrated within a $16\; \rm ns$ window.  The results are then 
sent from the camera's data-acquisition system to the control room via
optical fibres.

Soon after the second telescope became operational in January 2003, a
\lq Stereo' central trigger was implemented (June 2003), by which 
events are only retained if multiple telescopes see the same cascade.
This decreases the dead-time for the individual telescopes, allowing the
trigger threshold to be decreased (thus decreasing detector's energy
threshold), while the multiple images of each cascade provide a
increase in the background-rejection capability and the angular and
energy resolution of the system.

The Phase-I of \hess\ was completed in December, 2003, with the
addition of the fourth telescope, since which time the system has been
operating at its full sensitivity.  The energy threshold of the system
is $\sim 120\;\rm GeV$ for sources close to Zenith after background
rejection cuts ($\sim 400\;\rm GeV$ for single-telescope mode) with an
angular resolution improved to 0.06\dg\ (from 0.1\dg) and allowing
spectral measurements with an energy resolution of $\simeq 15\%$.
Observations of the Crab nebula have confirmed the system's
performance, with a rate of $10.8\;\rm \gamma/minute$ and a detection
significance of $26.6\;\sigma/\rm\surd hr$, which when extrapolated
for a sources close to Zenith give a 1 Crab-level sensitivity
($5\sigma$ detection) in only 30 seconds (1\% Crab in 25 hrs). See
\cite{HESS_WH,HESS_PV} for further details.

\section{Observations of AGNs with \bighess }

Since the first operation of the \hess\ detector, many galactic and
extragalactic sources have been studied.  The observation of AGNs at
the highest energies is a probe of the emission mechanisms in the jets
of these sources, and studies of the their multi-wavelength spectral
energy distributions (SEDs) and correlated variability over wavelength
enable emission models (leptonic or hadronic) to be tested.  In
addition, as these VHE photons interact with the intergalactic
Infra-Red (IIR) background (to give an electron-positron pair) and are
thus absorbed, they can also serve as a probe of this background
(resulting mainly from early star formation) which is difficult to
measure by direct means.  However, this absorption limits the distance
at which we can see AGNs to a redshift 0.5 at the \hess\ detector
threshold energy.  The large detection area of \hess\ allows us to
measure spectral and temporal characteristics on hour timescales
(depending on the strength of flares) for the sources seen.

Among the extra-galactic targets (with observing time up to Summer,
2004 in parentheses) are: \pks\ (92h), {\small PKS}$\;2005$-489 (52h),
M87 (32h), {\small NGC}253 (34h).  Here we present results from two
AGNs:
\pks\ and \mkn .

\subsection{The AGN \pks }

\pks\ is the brightest AGN in the Southern Hemisphere, and has been
well studied in many energy bands over the last 20 years.  It has been
previously detected at VHE energies \cite{durham}.  With a redshift of
$z=0.117$ it is one of the most distant VHE blazars, and therefore
of interest not only for studies of this class of object, but also for
IIR studies.

Initial observations were taken over all the installation phase of 
\hess\ Phase-I from July 2002 to October 2003, with an evolving 
detector threshold and sensitivity.  Clear detections ($>5\sigma$) are
seen in each night's observations, and an overall signal of
$44.9\sigma$ in $63.1\;\rm h$ of this mixed data, with $\sim
1.2\;\gamma\rm /min$, 10-60\% Crab level, with variability on
time-scales of months, days, and hours.  The energy spectra are
characterized by a steep power law with a time-averaged photon index
of $\alpha=-3.31\pm 0.06$.

Owing to a particularly high level seen by \hess\ in October, 2003,
we triggered our RXTE \lq\lq target of opportunity" proposal on this 
source, enabling quasi-simultaneous observations to be taken between
the two instruments.  Short-term variations ($<30\;\rm min$) are seen
in both these datasets, and multi-wavelength correlations will be
published in a forthcoming paper.

A \hess\ multi-wavelength campaign with the PCA instrument on board
the Rossi X-ray Timing Explorer (RXTE) has been successfully completed
in August, 2004, with the full four-telescope Phase-I array, and
therefore full sensitivity, and these data are under analysis.  This
intense study of this source should yield insights into its inner
workings.

\subsection{The AGN \mkn }

\mkn\ was the first extra-galactic source detected at VHE energies 
\cite{me}.
It is the closest 
such source (at $z=0.03$) and so is little affected by IIR absorption.
With a declination $\delta \sim 38^\circ$, it is still accessible to
\hess , though culminating at a Zenith angle above 60\dg .  Under
these conditions, observations with the \hess\ detector have a higher
threshold, but a compensatory larger effective area (as the light-pool
is geometrically larger for showers developing at a greater
atmospheric slant distance), and so gives access to the highest
energies of the spectrum.

In April of this year, a great increase in activity from this source
was seen by the all-sky monitor aboard RXTE, reaching an
historically-high level of $110\;\rm mCrab$ in mid-April.  A
multi-wavelength campaign was therefore triggered on this source,
including other IACTs, radio and optical telescopes, and RXTE.

The \hess\ observations, at an average Zenith angle of 62\dg, provided
a very clear signal in April, with $66\sigma$ in $9.71\;\rm h$ of
data, yielding $\sim 5.1\;\gamma\rm /min$, and an estimated 1-2 Crab
level.  The flux clearly increases from the January level ($6\sigma$
in $2.12\;\rm h$, $\sim 0.8\;\gamma\rm /min$, 10-50\% Crab level), and
was also seen by other IACTs in the Northern hemisphere (Whipple,
{\small MAGIC}).  Shorter-term variations and correlations with other
energy domains are currently under study.

\section{Future Plans for expansion to \bighess\ Phase-II}

Plans for Phase-II of the experiment are comprised of a large
telescope in the centre of the current Phase-I providing a lowered
threshold and increased sensitivity.  This will provide access to a
number of astrophysical phenomen\ae, such as the spectral cut-offs in
pulsars, microquasars, GRBs, and dark matter in the form of WIMPs.  As
concerns this paper, AGNs can be observed up to redshift of 2-3 with
\hess\ Phase-2 (vs. 0.5 with \hess ), provided that they are
sufficiently bright, as the optical depth due to absorption in the
intergalactic infra-red background is smaller at lower energies.
With detections of a larger number of AGNs at varying redshifts,
the effect of IIR absorption may be disentangled from the intrinsic
spectra of the sources.

Technical plans for this very-large telescope are well advanced. The
mount and dish structure (30\rm m \O) are well within the capabilities
of industry, since much larger radio-telescopes have been built. The
camera, using the same technology as Phase-I, with some improvements
in order to decrease the dead-time and readout speed, will have $\sim
2000$ pixels of size 0.05\dg ($\sim 3$\dg field of view).  An improved
Analogue Memory ASIC (Application-Specific Integrated Circuit) is
being prototyped, and the associated camera and read-out electronics
are being designed, based on the experience gained with the Phase-I.

In operation with the four telescopes of Phase-I, Monte Carlo
simulations indicate that, in coincidence mode the \lq 4+1' system
would have a detection threshold of $\sim 50\rm GeV$ with fine-grained
and photon-rich image in the central telescope providing improved
background rejection and angular and energy resolution.  In
stand-alone mode, a threshold as low as $15-25\; \rm GeV$ may be
achieved, though with lower background-rejection capability.

\section{Conclusions}

Phase-I of \hess\ has already provided many interesting new results,
of which some of those from extra-galactic sources are presented here.
Based on the experience gained with \hess\ Phase-I, a Phase-II
extension consisting of a very large Cherenkov Imaging Telescope is
being designed, which will provide an unprecedentedly low threshold
IACT, while greatly increasing the sensitivity at current energies.
\hess\ Phase-I will continue to provide exciting new results
in the future, while the Phase-II is being designed and installed.

\begin{figure}[b]
\begin{center}
\includegraphics[width=.99\textwidth]{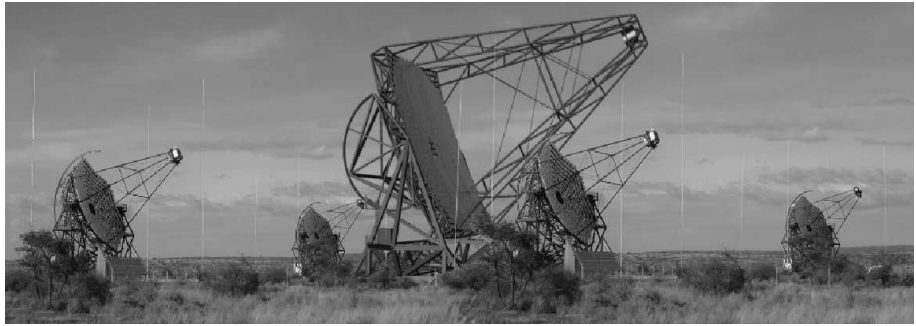}
\end{center}
\caption[]{Photo of the current four-telescope \hess\ Phase-I array,
with an artist's impression of the Phase-II 30m \O\ telescope in the
centre of the array superimposed.
}
\label{eps1}
\end{figure}

\end{document}